\magnification=\magstephalf
\input amstex
\loadbold
\documentstyle{amsppt}
\refstyle{A}
\NoBlackBoxes

\vsize=7.5in

\def\pf{\hfill $\square$}
\def\c{\cite}

\def\fg{\frak{g}}
\def\fh{\frak{h}}

\def\end{\text{End}}

\def\bl{\boldkey l}
\def\br{\boldkey r}

\topmatter
\title Coboundary dynamical Poisson groupoids and 
        integrable systems  \endtitle
\leftheadtext{L.-C. Li}
\rightheadtext{Poisson groupoids and integrable systems}

\author Luen-Chau Li\endauthor
\address{L.-C. Li, Department of Mathematics,Pennsylvania State University
University Park, PA  16802, USA}\endaddress
\email luenli\@math.psu.edu\endemail
\abstract  In this paper, we present a general scheme to construct
integrable systems based on realization in the coboundary dynamical
Poisson groupoids of Etingof and Varchenko.  We also present a
factorization method for solving the Hamiltonian flows.  To illustrate
our scheme and factorization theory, we consider a family of hyperbolic 
spin Ruijsenaars-Schneider
models related to affine Toda field theories and solve the equations of motion 
in a simple case.
\endabstract
\endtopmatter

\document
\subhead
1. \ Introduction
\endsubhead

\baselineskip 15pt
\bigskip

Many important examples of integrable systems are related to Lie groups
and Lie algebras and can be studied using a well-known group-theoretic
scheme (see, for example, \c{A}, \c{K}, \c{S}, \c{RSTS1}, \c{RSTS2}, \c{AvM},
\c{STS1},\c{STS2}, \c{RSTS3}, \c{FT}, \c{LP1} and the references therein).  
However, as the author has come to realize in the last few years, Lie 
algebroids and Lie groupoids are also of relevance \c{LX1},\c{LX2}, \c{L1}, 
\c{HM}.  More precisely, the requisite objects are connected with the
so-called classical dynamical r-matrices (or generalizations), which
first appeared in the context of Wess-Zumino-Witten (WZW) conformal
field theory \c{BDF},\c{F}.  While classical r-matrices play a role in
Poisson Lie group theory \c{D}, the authors in \c{EV} showed that an 
appropriate geometric setting for the classical
dynamical r-matrices is that of a special class of Poisson groupoids
(a notion due to  Weinstein  \c{W}), the so-called coboundary dynamical
Poisson groupoids.  If $R$ is an $H$-equivariant classical dynamical
r-matrix, and $(\Gamma, \{\,\cdot,\cdot \,\}_{R})$ is the associated
coboundary dynamical Poisson groupoid, then it follows from Weinstein's
coisotropic calculus \c{W} or otherwise that the Lie algebroid dual
$A^{*}\Gamma$ also has a natural Lie algebroid structure \c{LP2}, \c{BKS}.
We shall call $A^{*}\Gamma$ the coboundary dynamical Lie algebroid
associated to $R$ and it is this class of Lie algebroids which provides
the natural setting in \c{LX1}, \c{LX2}.  

At this point, it is convenient to summarize a few characteristics of the
class of invariant Hamiltonian systems which
admit either a realization in the dual vector bundle $A\Gamma$ of
$A^{*}\Gamma$ \c{LX1}, \c{LX2}, or in the coboundary dynamical Poisson
groupoid $\Gamma$ istself \c{L1}: (a) the systems are defined on a
Hamiltonian $H$-space $X$ with equivariant momentum map $J$ and the
Hamitonians are the pull-back of natural invariant functions by an
$H$-equivariant realization map, (b) the pullback of natural invariant
functions do not Poisson commute everywhere on $X$, but they do so
on a fiber $J^{-1} (\mu)$ of the momentum map, and (c) the reduced
Hamiltonian systems on $X_{\mu} = J^{-1} (\mu)/H_{\mu}$ ($H_{\mu}$ is the
isotropy subgroup at $\mu$) admit a natural collection of Poisson
commuting integrals.  

We should emphasize that while the reduced
systems are the goal, the unreduced ones are the key players in the
analysis.  Clearly, the geometric objects $\Gamma$ and $A\Gamma$ encode 
the hidden symmetries of the Hamiltonian systems and this is one of 
their virtues.   Now, while this is true for any
$\Gamma$ or $A\Gamma$ regardless of the underlying r-matrix, we shall
focus on those which correspond to the solutions of the modified
dynamical Yang-Baxter equation (mDYBE).  As the reader will see in
the present work, the (mDYBE) is connected with
a factorization problem on the trivial Lie groupoid $\Gamma$.  Moreover,
there is a prescription to integrate the Hamiltonian flows on 
$J^{-1} (\mu)$ explicitly based on this factorization.  Hence, we
can obtain the induced flows on $X_{\mu}$ by reduction.  Surely,
this is the most important aspect of our theory, and one in which
the criticality of the Lie groupoid $\Gamma$ is clearly demonstrated.

Our objective in this paper is to present some of the essentials
of the factorization theory which we mentioned above.  Since several 
constructions are more transparent for the coboundary dynamical Poisson 
groupoids, we have decided in this first publication to present our theory 
for this 
particular case (see \c{L2} and \c{L3} for the algebroid case). However, 
it is clear that several basic results (Prop.4.4, Thm.4.6 and 
Corollary 4.8) are 
common to both frameworks.  

The paper is organized as follows.  In Section 2 and 3, we present
a general scheme to construct integrable systems based on realization
in coboundary dynamical Poisson groupoids.  In the case of a constant
r-matrix, our theory here is a slight extension of the standard r-matrix 
framework for Lax equations on Lie groups \c{STS1}, \c{STS2} 
(see Theorem 2.5).
In Section 4, we discuss the algebraic and geometric structures associated 
with the (mDYBE),
leading up to a factorization method for solving the Hamiltonian flows.  As
an illustration of our theory in these sections, we introduce a
family of hyperbolic spin Ruijsenaars-Schneider (RS) models in Section 5 and
solve the equations of motion in a simple case.  
Remarkably, one of the models in the $SL(N + 1,\Bbb C)$ case
has been shown to govern the soliton dynamics of the so-called $A^{(1)}_N$
affine Toda field theory \c{BH}. We plan to study this family of spin
RS models as well as others in subsequent publications.

\bigskip
\bigskip
\noindent {\bf Acknowledgment.}  The author would like to thank Serge 
Parmentier for showing him the argument in the second half of the proof of 
Proposition 4.4 (a).  He is also grateful to Martin Schmoll who taught him a 
useful command in \TeX. Finally, he would like to thank the referee for 
useful suggestions to make the paper more readable. 
\bigskip
\bigskip
\subhead
2. \ Realization of Hamiltonian systems in coboundary dynamical Poisson 
\linebreak \phantom{fak}\, groupoids
\endsubhead

\bigskip
Let $G$ be a connected Lie group and $H\subset G$ a connected Lie subgroup.
We shall denote by $\fg$ and $\fh$ the Lie algebras corresponding to $G$
and $H$ respectively and let $\iota:\fh \longrightarrow \fg$ be the Lie
inclusion. 

We begin by recalling a fundamental construction of \c{EV} using the
formulation in \c{LP2}. Let $U\subset \fh^*$ be a connected 
$Ad_H^*$-invariant open subset, we say that a
smooth map $R:
U\longrightarrow L(\fg^*, \fg)$ (here and henceforth we denote by
$L(\fg^*,\fg)$ the set of linear maps from $\fg^*$ to $\fg$)  is a classical
dynamical $r$-matrix
if and only if  it is pointwise skew-symmetric:
$$<R(q)(A), B>=- <A, R (q) B>\eqno (2.1)$$
and satisfies the classical dynamical Yang-Baxter condition
$$\eqalign {&[R(q)A, R(q)B] +R(q)(ad^*_{R(q)A}B-ad^*_{R(q)B}A)\cr
+&dR(q)\iota^*A(B) - dR(q)\iota^*B(A) + d<R(A),B>(q) 
= \chi (A,B),\cr} \eqno (2.2)$$
where $\chi : \fg^*\times \fg^* \longrightarrow  \fg$ is
$G$-equivariant, that is,
$$ \chi (Ad^*_{g^{-1}}A, Ad^*_{g^{-1}}B) = Ad_g\, \chi(A,B)\eqno(2.3)$$
for all $A, B\in \fg^*$, $g\in G$, and all $q\in U$.
\smallskip
The dynamical $r$-matrix is said to be $H$-equivariant
if and only if
$$R(Ad^*_{h^{-1}} q)=Ad_h \circ R(q)\circ Ad^*_h \eqno (2.4)$$
for all $h\in H, q\in U$.
We shall equip $\Gamma=U\times G\times U$ with the trivial Lie groupoid
structure over $U$ \c{M} with target and source maps
$$\alpha (u,g,v)= u, \quad  \beta(u,g,v)= v \eqno(2.5)$$
and multiplication map
$$m((u,g,v), (v,g',w))= (u,gg',w).\eqno(2.6)$$

For $\varphi \in C^\infty (\Gamma)$, we define its partial derivatives
and its left and right gradients (with respect to $G$) by 
$$\eqalign {&<\delta_1 \varphi, u'>= {d\over dt}_{|_{t=0}} 
\varphi(u+tu', g, v),\quad <\delta_2 \varphi, v'>= {d\over dt}_{|_{t=0}} 
\varphi(u, g, v+tv'),\,
u', v'\in \fh^{*}\cr
&<D\varphi, X>= {d\over dt}_{|_{t=0}} \varphi(u, e^{tX}g,v),
\quad <D' \varphi, X>= {d\over dt}_{|_{t=0}} 
\varphi(u, ge^{tX}, v), \, X\in \fg.\cr}$$

\proclaim
{Theorem 2.1 \c{EV}} 
\smallskip
\noindent (a) The  bracket 
$$\eqalign {&\{\varphi,\psi\}_R (u,g,v)= <u, [\delta_1\varphi,\,
\delta_1\psi]> -<v, [\delta_2\varphi,\, \delta_2\psi]>\cr
&\hskip 50 pt - <\iota\delta_1 \varphi,\, D\psi> - <\iota\delta_2 \varphi, 
D'\psi>\cr
&\hskip 50 pt + <\iota\delta_1 \psi,\, D\varphi> + 
<\iota\delta_2 \psi , D'\varphi>\cr
&\hskip 50 pt +<R(v) D'\varphi, D'\psi> - <R(u) D\varphi, D\psi>\cr}\eqno(2.7)
$$
defines a Poisson structure on $\Gamma$ if and only if
$R: U\longrightarrow L(\fg^*, \fg)$ is an $H$-
equivariant dynamical $r$-matrix.
\smallskip
\noindent (b) The trivial Lie groupoid $\Gamma$ equipped with the Poisson 
bracket
$\{\, \cdot, \cdot \,\}_R$ is a Poisson groupoid.
\endproclaim
\smallskip
We shall call the pair $(\Gamma, \{\,\cdot, \cdot \,\}_R)$ the coboundary 
dynamical Poisson groupoid associated to $R$.

\definition
{Definition 2.2}  A Poisson manifold $(X, \{\,\cdot,\cdot \, \}_{X})$ is said 
to admit
a realization in the coboundary dynamical Poisson groupoid 
$(\Gamma, \{\,\cdot,\cdot \,\}_R)$ if there is a Poisson map
$\rho: (X,\{\,\cdot,\cdot \, \}_{X}) \longrightarrow (\Gamma, \{\,\cdot, \cdot \,\}_R)$.
\enddefinition

\definition
{Definition 2.3} Suppose a Poisson manifold $(X,\{\,\cdot,\cdot \,\}_{X})$ 
admits a 
realization $\rho: (X,\{\,\cdot,\cdot \,\}_{X}) \longrightarrow 
(\Gamma, \{\,\cdot,\cdot \,\}_R)$
and ${\Cal H}\in C^\infty (X)$.  The map $\rho$ is said to give
a realization of the Hamiltonian system
$\dot x = X_{\Cal H} (x)$ in $\Gamma$ if there exists
$\varphi\in C^\infty (\Gamma)$ such that ${\Cal H} = \rho^* \varphi$.
\enddefinition

In what follows, we shall work with a Poisson manifold 
$(X,\{\,\cdot,\cdot \,\}_{X})$ 
together with a realization $\rho: X\longrightarrow \Gamma$.  Let
$Pr_2: \Gamma \longrightarrow G$ be the projection map onto the
second factor of $\Gamma$ and set 

$$ L= Pr_2 \circ \rho : X \longrightarrow G, \eqno(2.8)$$
$$ m_1 = \alpha \circ \rho : X \longrightarrow U, \eqno(2.9)$$
\noindent and $$ m_2 = \beta \circ \rho : X \longrightarrow U,\eqno(2.10)$$
\noindent i.e. $\rho = (m_1, L, m_2)$.

\proclaim
{Proposition 2.4}  For all $f_{1}, f_{2} \in C^\infty(G)$,

$$
\aligned
       &\{L^*f_{1}, L^*f_{2} \}_X\, (x)
       \\
       = & \langle R(m_2(x))\,D'f_{1} (L(x)), D'f_{2} (L(x))\rangle
       \\
       & - \langle R(m_1(x))\,Df_{1} (L(x)), Df_{2} (L(x))\rangle,
       \qquad \forall x \in X, 
\endaligned
$$
\endproclaim

\demo
{Proof} Since $\rho$ is a Poisson map, we have

$$
\aligned
       &\{L^*f_{1}, L^*f_{2} \}_X\, (x)
       \\
       = & \{Pr_2^*f_{1}, Pr_2^*f_{2} \}_R \,(\rho (x))
       \\
       = & \langle R(m_2(x))\,D'f_{1} (L(x)), D'f_{2} (L(x))\rangle
       \\
       & - \langle R(m_1(x))\,Df_{1} (L(x)), Df_{2} (L(x))\rangle,
\endaligned
$$
as $\delta_i(Pr_2^*f_{1}) = \delta_i(Pr_2^*f_{2}) = 0$, $i=1,2$. \pf

\enddemo

Let $I(G)$ be the space of central functions on $G$. We now examine the class 
of Hamiltonian systems 
$\dot x = X_{\Cal H} (x)$ on $X$
which can be realized in $(\Gamma, \{\,\cdot,\cdot \,\}_R)$ by means of 
$\rho$ with
${\Cal H} \in \rho^*(Pr_2^*I(G))=L^*I(G)$.

\proclaim
{Theorem 2.5}
\smallskip
\noindent (a) If ${\Cal H}=L^*f$, where $f\in I(G)$, then under the flow
$\phi_t$ generated by the Hamiltonian $\Cal H$, the following equation 
hold:

$$
\aligned
       &{d\over dt} m_1(\phi_t)= -\iota^*\,Df(L(\phi_t)),
       \\
       &{d\over dt} L(\phi_t) = T_e r_{L(\phi_t)}\, R(m_1(\phi_t))
       (Df (L(\phi_t)))
       \\
       &\hskip 50pt -T_e l_{L(\phi_t)}\, R(m_2(\phi_t))(Df (L(\phi_t))),
       \\
       &{d\over dt} m_2(\phi_t)= -\iota^*\,Df(L(\phi_t)).
\endaligned
$$
\smallskip
\noindent (b) For all $f_1, f_2 \in I(G)$,
$$
\aligned
       & \{L^*f_1, L^*f_2\}_X\,(x)
       \\
       & = \langle (R(m_2(x))-R(m_1(x)))Df_1(L(x)), Df_2(L(x)) \rangle, 
       \qquad \forall  x \in X.
\endaligned
$$
\endproclaim

\demo
{Proof} (a) Since $\rho$ is a Poisson map, we have ${d\over dt}\rho(\phi_t)
= X_{Pr_2^*f} (\rho(\phi_t))$.  The equations then follow from a direct
computation using (2.7).
\newline
(b) This is obvious from Proposition 2.4.  \pf
\enddemo

Consider the important special case where $X=\Gamma$ and $\rho=id_{\Gamma}.$
If $R$ is a constant r-matrix, it is immediate from the above results
that the family of functions $Pr^{*}_{2} I(G)$ Poisson commutes on $\Gamma$.
So in this case, what we have here is a slight extension of the standard
r-matrix framework for Lax equations on Lie groups \c{STS1},\c{STS2}.
On the other hand, if $R$ is genuinely dynamical, the functions in
$Pr^{*}_{2} I(G)$ no longer Poisson commute on all of $\Gamma$, but
they do so on the gauge group bundle (see (3.3)), where 
$u=v$.  In this case, a reduction is required to obtain the associated
integrable systems.  In the rest of the paper, we shall deal 
exclusively with the genuinely dynamical case.

\bigskip
\bigskip
\subhead
3. \ Reduction to integrable systems
\endsubhead

\bigskip

Theorem 2.5 (b) shows that the functions in $L^*I(G)$ do not necessarily
Poisson commute in $X$.  In the following, we shall describe a situation
where we can obtain integrable flows on a reduced Poisson space.  In
particular, this applies to the important case where $X$ is $\Gamma$
itself and where $\rho$ is the identity map.  We shall make the following
assumptions:
\medskip

\noindent A1. $X$ is a Hamiltonian $H$-space with an equivariant momentum map
    $J: X \longrightarrow \fh^*$, \newline
\noindent A2. the realization map $\rho : X \longrightarrow \Gamma$ is $H$-
    equivariant, where $H$ acts on $\Gamma$ via the formula

$$h\cdot(u, g, v)=(Ad^*_{h^{-1}}u, hgh^{-1}, Ad^*_{h^{-1}}v). \eqno(3.1)$$

\noindent A3. for some regular value $\mu \in \fh^*$ of $J$,

$$\rho (J^{-1} (\mu)) \subset {\Cal I} \Gamma, \eqno(3.2)$$
\noindent where
$${\Cal I} \Gamma = \{ (u,g,u) \mid u \in U, g \in G \} \eqno(3.3)$$
\noindent is the gauge group bundle of $\Gamma.$

Now, recall that $\Gamma$ has a pair of commuting Hamitonian $H$-actions
\c{EV}:a left action given by $h\cdot(u,g,v)=(Ad^*_{h^{-1}}u, hg, v)$, 
and a right action given by $(u, g, v)\cdot h=(u, gh, Ad^*_h v)$.  By
combining these actions, we obtain the $H$-action in assumption A2.  As
can be easily verified, this is a Hamiltonian action with equivariant
momentum map

$$\gamma = \alpha - \beta: \Gamma \longrightarrow \fh^* \eqno(3.4)$$
and we have $\gamma^{-1} (0) = {\Cal I} \Gamma.$  Consequently,
$X=\Gamma$ and $\rho=id_{\Gamma}$ satisfy assumptions A1-A3 with
$\mu = 0.$

We shall denote by $H_{\mu}$ the isotropy subgroup of $\mu$ for the
$H$-action on $X$.  Then it follows by Poisson reduction \c{MR},
\c{OR} (see \c{OR} for the singular case) that the variety
$X_{\mu} = J^{-1}(\mu)/ H_{\mu}$ inherits a unique Poisson structure 
$\{\,\cdot,\cdot \,\}_{X_{\mu}}$
satisfying 

$$\pi_{\mu}^* \{f_1, f_2 \}_{X_{\mu}} = i_{\mu}^* \{\widetilde{f_1},
\widetilde{f_2} \}_X .\eqno(3.5)$$
Here, $i_{\mu}:J^{-1} (\mu) \longrightarrow X$ is the inclusion map,
$\pi_{\mu}:J^{-1} (\mu)\longrightarrow X_{\mu}$ is the canonical projection,
$f_1$, $f_2 \in C^{\infty} (X_{\mu})$, and $\widetilde{f_1}$,
$\widetilde{f_2}$ are (locally defined) smooth extensions of 
$\pi_{\mu}^* f_1$, $\pi_{\mu}^* f_2$ with differentials vanishing
on the tangent spaces of the $H$-orbits.  For the case where
$X=\Gamma$, $\rho=id_{\Gamma}$, it is clear that the isotropy
subgroup $H_0$ of $\mu = 0$ is $H$ and so we have the reduced Poisson
variety

$$(\Gamma_0 = {\gamma^{-1}(0)/ H}, \{\,\cdot,\cdot \,\}_{\Gamma_0}),
\eqno(3.6)$$
with 
the inclusion map $i_H  :\gamma^{-1}(0) \longrightarrow \Gamma$ and
the canonical projection $\pi_H :\gamma^{-1}(0) \longrightarrow \Gamma_0.$

Clearly, functions in $i_H^* Pr_2^* I(G) \subset C^{\infty} (\gamma^{-1}(0))$
are $H$-invariant, hence they descend to functions in 
$C^{\infty} (\Gamma_0)$.  On the other hand, it follows from
assmption A2 that the functions in $i_{\mu}^* L^* I(G) \subset
C^{\infty} (J^{-1}(\mu))$ drop down to functions in $C^{\infty} (X_{\mu}).$  Now,
by assumptions A2-A3, and the fact that $\rho$ is Poisson, it follows
from \c{OR} that $\rho$ induces a unique Poisson map

$$\widehat{\rho}: X_{\mu} \longrightarrow \Gamma_0 = {{\Cal I}\Gamma
/ H} \eqno(3.7)$$
characterized by  $\pi_H \circ \rho \circ i_{\mu}=\widehat{\rho}
\circ \pi_{\mu}.$  Hence $X_{\mu}$ admits a realization in the
Poisson variety $\Gamma_0$.

We shall use the following notation.  For $f \in I(G)$, the unique
function in $C^{\infty} (\Gamma_0)$ determined by $i_H^* Pr_2^* f$
will be denoted by $\bar f$; while the unique function in
$C^{\infty} (X_{\mu})$ deetermined by $i_{\mu}^* L^* f$ will be
denoted by ${\Cal F}_{\mu}$.  From the definitions, we have

$${\Cal F}_{\mu} \circ \pi_{\mu} = ({\widehat{\rho}}^{*} \bar f)
  \circ \pi_{\mu} = i_{\mu}^{*}L^{*} f.\eqno(3.8)$$

\proclaim
{Theorem 3.1} Let $(X, \{\,\cdot,\cdot \,\}_X )$ be a Poisson manifold which
admits a realization $\rho: X \longrightarrow \Gamma$ and assume that
A1-A3 are satisfied.  Then there exist a unique Poisson structure
$\{\,\cdot,\cdot \,\}_{X_{\mu}}$ on the reduced space $X_{\mu} = 
J^{-1}(\mu)/H_{\mu}$
and a unique Poisson map $\widehat{\rho}$ such that

\noindent (a) for all $f_1$, $f_{2}\in I(G)$ and $x\in J^{-1} (\mu)$,

$$
\aligned
       & \{{\widehat{\rho}}^{*}{\bar f}_1, {\widehat{\rho}}^{*}
         {\bar f}_2 \}_{X_{\mu}}\circ \pi_{\mu} (x) \\
     =\,& <R(m_{1}(x)) D'f_{1}(L(x)), D'f_{2}(L(x))>\\
        & - <R(m_{1}(x)) Df_{1}(L(x)), Df_{2} (L(x))>,\\
\endaligned
$$
\smallskip
\noindent (b) functions ${\widehat{\rho}}^{*}{\bar f}$, $f\in I(G)$, 
Poisson commute in $(X_{\mu}, \{\,\cdot,\cdot \,\}_{X_{\mu}})$,
\smallskip
\noindent (c) if $\psi_t$ is the induced flow on ${\Cal I}\Gamma$ 
generated by the Hamiltonian $Pr^*_{2} f$, $f \in I(G)$, and 
$\phi_t$ is the Hamiltonian flow of ${\Cal F}=L^{*} f$ on $X$, then
the reduction $\phi^{red}_t$ of $\phi_{t} \circ i_{\mu}$ on
$X_{\mu}$ defined by $\phi^{red}_t \circ \pi_{\mu} = \pi_{\mu} \circ
\phi_{t} \circ i_{\mu}$ is a Hamiltonian flow of ${\Cal F}_{\mu} =
{\widehat {\rho}}^{*} {\bar f}$ and ${\widehat {\rho}}\circ \phi^{red}_{t}
(\pi_{\mu} (x)) = \pi_{H} \circ \psi_{t} (\rho(x))$, \quad $x \in J^{-1}
(\mu)$.
\endproclaim

\demo
{Proof} (a) Since $\rho (J^{-1} (\mu)) \subset {\Cal I} \Gamma$, we
have $m_{1} (x)=m_{2}(x)$ for $x \in J^{-1} (\mu)$.  Therefore,

$$
\aligned
       &\{{\widehat{\rho}}^{*}{\bar f}_1, {\widehat{\rho}}^{*}
         {\bar f}_2 \}_{X_{\mu}}\circ \pi_{\mu} (x) \\
     =\,& \{{\bar f}_1, {\overline f}_{2} \}_{\Gamma_0} \circ \pi_{H}
         (\rho (x)) \\
     =\,&\{Pr^{*}_{2} f_1, Pr^{*}_{2} f_2 \}_{R}
         (\rho (x)) \\
\endaligned
$$
and so the assertion now follows from the proof of Proposition 2.4.
\newline
(b) This assertion is clear from part (a).
\newline
(c) The first part of the assertion follows from Theorem 2.16 of 
\c{OR} while the second part is a consequence of the
relation $\rho \circ \phi_{t} \circ i_{\mu} = \psi_{t} \circ \rho \circ
i_{\mu}$ and the definition of ${\widehat \rho}$ and $\phi^{red}_t$.
\pf
\enddemo

\noindent {\bf Remark 3.2} (a) If we assume the existence of an
$H$-equivariant map $g:X \longrightarrow H$, then we can define
the gauge transformation ${\widetilde \rho}$ of the realization map $\rho$
in the obvious way.  Therefore, if the
reduced Poisson space $X_{\mu}$ is smooth (or has a smooth
component), then ${\widetilde \rho}\big|_
{J^{-1}(\mu)}$ descends to a uniquely determined map $\rho_{\mu}$ on
$X_{\mu}$ and we can write down the (generalized) Lax equations for
the the reduced Hamiltonian systems on
$X_{\mu}$ (or a smooth component of $X_{\mu}$).  
Details of this has been work out in \c{L1} but we
do not need it in this work.
\smallskip
\noindent (b) Clearly, our results here can be generalized in several
directions.  For example, we can consider coboundary Poisson groupoids
$(M\times G\times M, \{\,\cdot,\cdot \,\})$ (where $M$ is a manifold)
corresponding to the Lie bialgebroids considered in \c{LiX2}.  Another
possibility is to consider realization in $\Gamma^{N}=\Gamma\times
\cdots\times \Gamma$ ($N$ times), equipped with the product structure,
and use the class of twisted invariant
functions on $G^{N}$.  We shall report on these results elsewhere.
\smallskip
\noindent (c) An example of integrable systems which fits into our
framework is given by the elliptic Sklyanin systems in \c{HM}.  
Therefore, the Hamilton's equations of such systems can be solved
via our factorization theory in the next section (see \c{L3}).
In Section 5 below, we shall present an example of physical origin.

\bigskip
\bigskip

\subhead
4. \ Exact solvability and factorization problems on Lie groupoids
\endsubhead

\bigskip

It is well-known that an important class of Poisson Lie groups is
associated with the Sklyanin bracket \c{D},\c{STS2}.  For Poisson groupoids,
an analog of the Sklyanin construction was considered in \c{LiX1}.  We
shall begin by relating the bracket $\{\,\cdot,\cdot \,\}_R$ in Theorem 2.1 to
the consideration in \c{LiX1}.  Let $A \Gamma := \bigcup_{u \in U} T_
{\epsilon(u)} \alpha ^{-1} (u) = \bigcup_{u\in U} \lbrace 0_{u} \rbrace
\times \fg \times \fh^*$ be the Lie algebroid of $\Gamma.$    By Weinstein's
coisotropic calculus \c{W} or otherwise, the Lie algebroid dual $A^*\Gamma
= \bigcup_{u\in U} \lbrace 0_{u} \rbrace \times \fg^* \times \fh$ also
has a natural Lie algebroid structure \c{BKS},\c{LP2}  such that the pair
$(A\Gamma, A^*\Gamma)$ is a Lie bialgebroid in the sense of Mackenzie
and Xu \c{MX}.  We shall denote the Lie brackets on $Sect(U,A\Gamma)$
and $Sect (U,A^*\Gamma)$, respectively, by $[\cdot, \cdot]_{A\Gamma}$
and $[\cdot, \cdot]_{A^*\Gamma}.$

For our purpose, we introduce the bundle map
$${\Cal R}: A^*\Gamma \longrightarrow A\Gamma, 
(0_{q}, A, Z)\mapsto (0_{q}, -\iota Z+R(q)A, \iota^*A-ad^*_{Z}q).
\eqno(4.1)$$

For $\varphi \in C^{\infty}(\Gamma)$, we also define the left and
right gradients ${\Cal D}'\varphi (u,g,v)$ and ${\Cal D}\varphi (u,g,v)$ as 
follows:

$$<{\Cal D}'\varphi (u,g,v), X(v)>={d\over dt}_{|_{t=0}} \varphi (r_{e^{tX}} 
(u,g,v)),
\eqno(4.2a)$$

$$<{\Cal D}\varphi (u,g,v), X(u)> ={d\over dt}_{|_{t=0}} \varphi (l_{e^{tX}} 
(u,g,v)),
\eqno(4.2b)$$
where $X \in Sect(U,A\Gamma)$, and $r_{e^{tX}}$, $l_{e^{tX}}$ are the right
and left translations corresponding to the local bisection $e^{tX}$,
defined by $r_{e^{tX}} (u,g,v) = (u,g,v) e^{tX} (v)$ and
$l_{e^{tX}} (u,g,v) = e^{tX} ((\beta \circ e^{tX})^{-1}(u)) (u,g,v)$,
respectively, and where $e^{tX}(u)= f_{t}(u,1,u)$, $f_{t}$ being the 
(local) flow generated by the left invariant vector field 
$\overleftarrow {X}$ (see [M] for more details on the exponential map, 
bisections and
the corr. left and right translations).

\proclaim
{Proposition 4.1}  For all $\varphi, \psi \in C^{\infty}(\Gamma)$, 

$$\eqalign {\{\varphi, \psi\}_R (u,g,v)
&=<{\Cal R}_v ({\Cal D}' \varphi (u,g,v)), {\Cal D}'\psi (u,g,v)>\cr
&-<{\Cal R}_u ({\Cal D} \varphi (u,g,v)), {\Cal D}\psi (u,g,v)>.\cr}
$$
\endproclaim

\demo
{Proof} From the definition of the left and right gradients, we can
check that ${\Cal D}' \varphi (u,g,v)=(0_v, D' \varphi, {\delta_2}\varphi)$
and ${\Cal D} \varphi (u,g,v)=(0_u, D \varphi, -{\delta_1}\varphi).$  The
rest of the proof is plain.  \pf
\enddemo

If $(u,g,v)\in \Gamma$, we shall denote by 
${\bl}_{(u,g,v)}$ and ${\br}_{(u,g,v)}$ the left 
translation
and right translation in $\Gamma$ defined by 
${\bl}_{(u,g,v)} (v,g^{\prime},w)=(u,g g^{\prime},w)$ and
${\br}_{(u,g,v)} (w,g^{\prime},u)=(w,g^{\prime} g, v)$ respectively. 

\proclaim
{Corollary 4.2}  The Hamiltonian vector field generated by $\psi \in
C^{\infty} (\Gamma)$ is given by

$$\eqalign {X_{\psi} (u,g,v) =&-T_{\epsilon(v)} {\bl}_{(u,g,v)} {\Cal R}_v 
({\Cal D}'\psi (u,g,v))
\cr
&-T_{\epsilon(u)} ({\br}_{(u,g,v)}\circ i) {\Cal R}_u ({\Cal D}\psi (u,g,v)),
\cr}$$
where $\epsilon$ is the identity section and $i$ is the inversion map
of $\Gamma$. 
\endproclaim

\demo
{Proof}  This proof follows from Proposition 4.1 and the relations

$$<{\Cal D}'\varphi (u,g,v), X(v)>=\overleftarrow {X} \varphi (u,g,v),$$
$$<{\Cal D} \varphi (u,g,v), X(u)>=\overrightarrow {X} \varphi (u,g,v),$$
where $\overleftarrow {X}$ (resp., $\overrightarrow {X}$) is the left
(resp., right) invariant vector field on $\Gamma$ generated by
$X \in Sect (U, A\Gamma)$.  \pf
\enddemo

\definition
{Definition 4.3} The bundle map ${\Cal R}: A^*\Gamma \longrightarrow A\Gamma$
is called an r-matrix of the Lie algebroid $A^{*}\Gamma$.
\enddefinition

An important sufficient condition for $\{\,\cdot,\cdot \,\}_R$ to be a 
Poisson bracket
is the modified dynamical Yang-Baxter equation (mDYBE):

$$\eqalign {& [R(q)A, R(q)B]+R(q)(ad^*_{R(q)A}B-ad^*_{R(q)B}A)\cr
+\,&dR(q)\iota^*A(B) - dR(q)\iota^*B(A) + d<R(A),B>(q)\cr
=\,& - [K(A), K(B)]\cr}\eqno(4.3)$$
where $K\in L(\fg^*,\fg)$ is a nonzero symmetric map which
satisfies $ad_X\circ K + K\circ ad^*_X = 0$ for all $X \in \fg$.

Using $K$, we define

$${\Cal K}:A^*\Gamma \longrightarrow A\Gamma, 
(0_q,A,Z) \mapsto (0_q,K(A),0),\eqno(4.4)$$
and set ${\Cal R}^{\pm}={\Cal R}\pm {\Cal K}$, $R^{\pm}(q) =R(q)\pm K.$

We shall give a sketch of the proof of our next result, details can be
found in \c{L2}.

\proclaim
{Proposition 4.4} If $R$ satifies the (mDYBE), then
\smallskip
\noindent (a) ${\Cal R}^{\pm}$ are morphisms of transitive Lie algebroids.  
In particular,
$$[{\Cal R}^{\pm}(0,A,Z), {\Cal R}^{\pm}(0,A',Z')]_{A\Gamma}
={\Cal R}^{\pm}[(0,A,Z),(0,A',Z')]_{A^*\Gamma} \eqno(4.5)$$
for all smooth maps $A, A':U \longrightarrow \fg^*$,
$Z, Z':U \longrightarrow \fh,$
\smallskip
\noindent (b) $Im {\Cal R}^{\pm}$ are transitive Lie subalgebroids of 
$A\Gamma$.
\endproclaim

\demo
{Proof} (a) First of all, we can show that the r-matrix ${\Cal R}$ satifies
the equation
$$\eqalign {& [{\Cal R}(0,A,Z),{\Cal R}(0,A',Z')]_{A\Gamma}-
{\Cal R}[(0,A,Z),(0,A',Z')]_{A^*\Gamma}\cr
=&(0,-[K(A),K(A')],0)\cr}\eqno(4.6)$$
for all smooth maps $A, A':U \longrightarrow \fg^*$,
$Z, Z':U \longrightarrow \fh.$
If $a$ and $a_*$ are the anchor maps of the Lie algebroids
$A\Gamma$ and $A^*\Gamma$, it is easy to check that $a\circ {\Cal R}^{\pm}
=a_*$.  On the other hand, it follows from (4.6) that (4.5)
holds if and only if

$$
\aligned
       & {\Cal K} [(0,A,Z),(0,A',Z')]_{A^*\Gamma}
       \\
     =\,& [{\Cal R}(0,A,Z), {\Cal K}(0,A',Z')]_{A\Gamma}
         + [{\Cal K}(0,A,Z), {\Cal R}(0,A',Z')]_{A\Gamma}
\endaligned
$$
This latter relation can then be verified in a direct manner by using 
the property that $ad_X\circ K + K\circ ad^*_X = 0$ for $X \in \fg$.
\newline
(b) This property is a consequence of (a).
\pf
\enddemo

\noindent {\bf Remark 4.5.}  An upshot of Proposition 4.4 (a) is that
the dual maps $({\Cal R}^{\pm})^*= -{\Cal R}^{\mp}$ are Poisson maps,
when the domain and target are equipped with the corresponding 
Lie-Poisson structures.  This fact has been used in \c{L2}to construct a 
family of hyperbolic spin Calogero-Moser systems.  In particular, the
associated integrable models contain as a special case the system considered
in \c{R}.

\smallskip

In the rest of the section, we shall assume that there is an ad-invariant 
nondegenerate pairing $(\cdot, \cdot)$ on $\fg$ and without loss of
generality, we shall take the map $K:\fg^* \longrightarrow \fg$ in the above
discussion to be the identification map induced by $(\cdot, \cdot)$.
Indeed, with the identification $\fg^* \simeq \fg$, we shall regard
$R$ as taking values in $End (\fg)$, and the left and right gradients
as well as the dual maps are computed using $(\cdot, \cdot)$.  Also,
we have $ad^* \simeq -ad$ and $\iota^* \simeq \Pi_{\fh}$, where $\Pi_{\fh}$
is the projection map to $\fh$ relative to the direct sum decomposition
$\fg = \fh \oplus \fh^{\perp}$.  We shall keep, however, the notation
$A^*\Gamma$ although as a set it can be identified with $A\Gamma.$

The connection between (mDYBE) and our factorization theory is the following
decomposition
$$\eqalign {(0_q,X,0)
             &={1 \over 2}{\Cal R}_q^+ (0_q,X,0)\cr
             &-{1 \over 2}{\Cal R}_q^- (0_q,X,0)\cr}\eqno(4.7)
$$
where the element $(0_q,X,0)$ on the left hand side is in 
the adjoint bundle $Ker\,a$ of $A\Gamma.$
The reader should 
note that  the vector bundles
$\bigl \{{\Cal R}_q^{\pm} (0_q,X,0)\mid q\in U,\,X\in \fg \bigr \}$
are not Lie subalgebroids of $A\Gamma$ unless $R$ is
a constant r-matrix (see Remark 4.11 (c)).  This fact has repercussion
when we try to formulate a global version of the decomposition in (4.7).
 
In order to state our next result, introduce the Lie algebroid direct sum 
$A\Gamma {\underset TU \to \oplus} A\Gamma.$
Clearly, this is the Lie algebroid of the product groupoid
$P = \Gamma {\underset U\times U\to \times} \Gamma \rightrightarrows U$ 
($\simeq$ the trivial Lie groupoid $U \times (G \times G)\times U$).
For later usage, we shall denote the structure maps (target, source etc.)
of $P$ by $\alpha_{P}$, $\beta_{P}$, and so forth.

\proclaim
{Theorem 4.6} The map $({\Cal R}^{+}, {\Cal R}^{-}):A^{*}\Gamma
\longrightarrow A\Gamma {\underset TU \to \oplus} A\Gamma$ is a monomorphism
of transitve Lie algebroids.  In particular, the coboundary dynamical Lie
algebroid $(A^{*}\Gamma, [\cdot, \cdot]_{A^{*}\Gamma})$ is integrable.
\endproclaim

\demo
{Proof} Since ${\Cal R}_q^+ (0_q,X,Z)-{\Cal R}_q^- (0_q,X,Z)=
(0_q,2X,0)$, it is clear that $({\Cal R}^{+}, {\Cal R}^{-})$ is 1:1.
On the other hand, it is consequence of Proposition 4.4 (a) and the
definition of $A\Gamma {\underset TU \to \oplus} A\Gamma$ that 
$({\Cal R}^{+}, {\Cal R}^{-})$ is a morphism of Lie algebroids.
Finally, as $A^{*}\Gamma$ is isomorphic to a Lie subalgebroid of
$A\Gamma {\underset TU \to \oplus} A\Gamma$ and the latter is
integrable, it follows that $A^{*}\Gamma$ is also integrable \c{MM}.
\pf
\enddemo

\noindent{\bf Remark 4.7} (a) Note that in contrast to the case of
finite dimensional Lie algebras, not all Lie algebroids are integrable
even in finite dimensions \c{AM}.  In Theorem 4.6, the
integrability of the Lie algebroid $A\Gamma {\underset TU \to \oplus} A\Gamma$
has essentially allowed us to bypass the necessity of checking the 
integrability conditions in \c{CF}. 
\smallskip
\noindent (b) In addition to its application below in our factorization
theory, this result also implies the existence of the dual Poisson
groupoids for the the class of coboundary dynamical Poisson groupoids
$(\Gamma, \{\,\cdot, \cdot \,\}_{R})$, where $R$ is a solution
of (mDYBE).
\smallskip
\noindent (c) From the proof of Theorem 4.6, it is clear that
$({\Cal R}^{+}, {\Cal R}^{-})(0_q,X,Z)$ is an element of the
diagonal of $A\Gamma {\underset TU \to \oplus} A\Gamma$ if and
only if $(0_q,X,Z) = (0_q,0,Z)$.  We shall use the global version
of this fact in Corollary 4.8 below.  
\medskip
In the rest of the section, we shall assume both $G$ and $U$ are simply-
connected. Let $\Gamma^*$ be the unique source-simply connected Lie groupoid
which integrates $(A^{*}\Gamma, [\cdot, \cdot]_{A^{*}\Gamma})$.  
Then $({\Cal R}^{+}, {\Cal R}^{-})$ can be lifted up to a unique
monomorphism of Lie groupoids $\Gamma^{*}:\longrightarrow\Gamma
{\underset U \times U \to \times}\Gamma$ which we shall denote
by the same symbol.  Now, let $j: \Gamma {\underset U \times U \to \times}
\Gamma \longrightarrow {\Cal I}\Gamma$ be the map defined by
$j(a,b) = ab^{-1}$ and let ${\widetilde  m} = j \circ 
({\Cal R}^{+}, {\Cal R}^{-})$.

Our next corollary is a global version of the decomposition in 
(4.7).  For its formulation, note that the Lie groupoid
of $\{(0_{q}, 0, Z)\mid q\in U, Z\in \fh\}\subset A^{*}\Gamma$ is
$H \times U$, with target and source maps $\alpha^{\prime} (h,u) =u$,
$\beta^{\prime} (h,u)= Ad_{h} u$ and multiplication map
$m^{\prime} ((h,u),(k,Ad_{h} u))=(kh,u)$ (this is isomorphic
to the Hamiltonian unit in \c{LP2}).  On the other hand, the Lie groupoid of
${\Cal R}^{\pm} \bigl \{(0_{q},0,Z)\mid q\in U, Z\in \fh \bigr \}$
is given by  $E =\{\,(u, h, Ad_{h^{-1}}u) \mid u\in U,\,
h\in H\,\}$ and ${\Cal R}^{\pm}$ embeds $H\times U$ in $E$,
${\Cal R}^{\pm}\mid H\times U: (h,u)\mapsto (u, h^{-1}, Ad_{h} u).$ 
Clearly, the diagonal $\Delta(E)$ of  
$E {\underset U \times U \to \times} E$ acts
on $Im ({\Cal R}^{+}, {\Cal R}^{-})$ from the right via the
simple formula 
$$
\aligned
       &((u, k_{+}, v),(u, k_{-},v))\cdot((v,h,Ad_{h^{-1}}v),(v, h, Ad_{h^{-1}}v))
        \\ 
   =& ((u, k_{+} h, Ad_{h^{-1}}v), (u,k_{-} h,Ad_{h^{-1}}v))
\endaligned
$$ 
and the map
$j \mid Im ({\Cal R}^{+}, {\Cal R}^{-})$ is constant on the orbits
of this action.

\proclaim
{Corollary 4.8} Suppose $U$ is simply-connected, then 
$j \mid Im ({\Cal R}^{+}, {\Cal R}^{-})$
induces a one-to-one map
${\widehat j}: {Im ({\Cal R}^{+}, {\Cal R}^{-})}/
\Delta(E) \longrightarrow
{\Cal I}\Gamma$.  Therefore, for each $\gamma \in Im \,{\widetilde m}$,
there exists a unique $[\,(\gamma_+,\gamma_-)\,]$ in the homogeneous
space ${Im ({\Cal R}^{+}, {\Cal R}^{-})} / \Delta(E)$
such that ${\widehat j} ([\,(\gamma_+,\gamma_-)\,]) =\gamma$.
\endproclaim

\demo
{Proof} Suppose $[((u, k_+, v), (u, k_-, v))]$ and 
$[((u, k'_+, v'), (u, k'_-, v'))]$ are two elements in
${Im ({\Cal R}^{+}, {\Cal R}^{-})}/ \Delta(E)$
with the same image under the map ${\widehat j}$, then
$k_{+}{k_{-}^{-1}} = k'_{+}{k'_{-}}^{-1}$ and therefore,
${k_{+}^{-1}}k'_{+} = {k_{-}^{-1}}k'_{-}.$  Now, 

$$
\aligned
       &((v,k_{+}^{-1}k'_{+},v'), (v,k_{-}^{-1}k'_{-},v')) \\
    =\,&((u,k_{+},v),(u,k_{-},v))^{-1} ((u,k'_{+},v'), (u,k'_{-},v'))\\
    =\,&({\Cal R}^{+} (\gamma^*), {\Cal R}^{-} (\gamma^*))\\
\endaligned
$$
for unique $\gamma^{*} \in \Gamma^*$.  Hence from the fact that
${\Cal R}^{+} (\gamma^*)={\Cal R}^{-} (\gamma^*)$, we must have 
$\gamma^*\in H\times U$ (from Remark 4.7 (c) and the above discussion) and so 
$k_{+}^{-1}k'_{+} = k_{-}^{-1}k'_{-}
=h \in H$ and $v' = Ad_{h^{-1}} v$.  Consequently, 
$[((u, k_+, v), (u, k_-, v))]$ = 
$[((u, k'_+, v'), (u, k'_-, v'))]$.
\pf
\enddemo
\smallskip
In what follows, we shall rescale the Poisson bracket 
$\{\,\cdot,\cdot\, \}_R$ by the factor $1/2$.  
Let $f\in I(G)$ and let $F=Pr^*_{2} f$.  Then
${\Cal D}'F(u,g,v)=(0_v,Df(g),0)$ and ${\Cal D}F(u,g,v)=(0_u,Df(g),0)$.  
Therefore,
according to Corollary 4.2, the Hamilton's equation generated by
$F$ when restricted to ${\Cal I}\Gamma$ takes the form

$$\eqalign {&{d \over dt} (u,g,u)\cr
         =\,& -{1 \over 2} T_{\epsilon(u)}{\bl}_{(u,g,u)} {\Cal R}_u
             (0_u,Df(g),0)\cr
            & -{1 \over 2} T_{\epsilon(u)} ({\br}_{(u,g,u)}\circ i) {\Cal R}_u
             (0_u,Df(g),0).\cr}\eqno(4.8)
$$

\proclaim
{Theorem 4.9}  Suppose that $f \in I(G)$,  $F=Pr_2^*f$ and $u_0 \in
U,$ where $U$ is simply-connected. Then for some 
$0 < T \leq \infty$, there exists a unique element 
$(\gamma_{+}(t),\gamma_{-}(t))=
((u_0, k_{+} (t), u(t)), (u_0, k_{-} (t), u(t))) \in
Im ({\Cal R}^{+}, {\Cal R}^{-})$
for $0 \leq t <T$ which is smooth in t, solves the factorization
problem

$$exp \{-t(0,Df(g_0),0)\}(u_0)
      =\,\gamma_{+} (t)\,\gamma_{-} (t)^{-1}\eqno(4.9)
$$
and satisfies 
$$\eqalign {(T_{\gamma_{+} (t)} {\bl}_{{\gamma_{+}(t)}^{-1}} {\dot\gamma_{+}(t)},
T_{\gamma_{-} (t)}{\bl}_{{\gamma_{-}(t)}^{-1}}{\dot \gamma_{-}(t)})
\in \, 
({\Cal R}^+,{\Cal R}^-)_{u(t)} (\{0_{u(t)}\}\times\fg\times \{0\})\cr}
\eqno(4.10a)$$
with
$$\gamma_{\pm}(0) = (u_{0},1,u_{0}).\eqno(4.10b)$$

\noindent Moreover, the solution of Eqn.(4.8) with initial data $(u,g,u)(0)=
(u_0,g_0,u_0)$ is given by the formula

$$\eqalign {&(u,g,u)(t)\cr
=\,&(u_0, k_{\pm} (t), u(t))^{-1}(u_0,g_0,u_0)(u_0, k_{\pm} (t), u(t))\cr}
\eqno(4.11)$$
\endproclaim

\demo
{Proof}  We first prove the uniqueness of the element 
$(\gamma_{+} (t),\gamma_{-}(t))$.  Suppose 
$(\gamma'_{+}(t), \gamma'_{-}(t))=
((u_0, k'_{+} (t), u'(t)), (u_0, k'_{-} (t), u'(t)))\in
Im ({\Cal R}^{+},{\Cal R}^{-})$
is a second element with the properties in (4.9) and (4.10).  Then 
by Corollary 4.8, we have ${k_{+}^{-1}}(t)k'_{+}(t) = {k_{-}^{-1}}(t)k'_{-}
(t)=h(t) \in H$ and $u'(t)= Ad_{h(t)^{-1}} u(t).$  Consider
$h(t) ={k_{+}^{-1}}(t)k'_{+}(t)$.  By differentiation, and using
(4.10), we have

$$\eqalign{& T_{h(t)} l_{h(t)^{-1}} {\dot h(t)}\cr
= \, & T_{k'_{+}(t)} l_{k'_{+}(t)^{-1}} {\dot k'_{+}(t)}
-Ad_{h(t)^{-1}} T_{k_{+}(t)} l_{k_{+}(t)^{-1}} {\dot k_{+}(t)}\cr
= \, & R^{+}(u'(t)) X'(t) - Ad_{h(t)^{-1}} R^{+}(u(t)) X(t)\cr}
\eqno(4.12)
$$
for some $X(t), X'(t)\in \fg.$  Similarly, by taking
$h(t) =k_{-}^{-1}(t)k'_{-}(t)$, we obtain
$$\eqalign{& T_{h(t)} l_{h(t)^{-1}} \dot h(t)\cr
= \, &R^{-}(u'(t)) X'(t) - Ad_{h(t)^{-1}} R^{-}(u(t)) X(t).\cr}
\eqno(4.13)
$$
Therefore, upon equating the two expressions, we find 
$X'(t)=Ad_{h(t)^{-1}} X(t)$.  Substituting this back in (4.12),
and using (2.4) and the relation  $u'(t)= Ad_{h(t)^{-1}} u(t)$,
we learn that $T_{h(t)} l_{h(t)^{-1}} \dot h(t)=0$.  Therefore,
$h(t)=1$ and so 
$(\gamma_{+}(t),\gamma_{-}(t))=(\gamma'_{+}(t),\gamma'_{-}(t))$.

Assuming the existence of the factors for the moment,
we claim that $(u(t),g(t),u(t))$ as given by (4.11) solves
(4.8).  First of all, we have
$$
\aligned
       & \gamma_{+}(t)^{-1} (u_0,g_0,u_0)\, \gamma_{+} (t) \\
    =\,& (u(t), k_{+}(t)^{-1} g_0 k_{+}(t), u(t)) \\
    =\,& (u(t), k_{-}(t)^{-1} g_0 k_{-}(t), u(t)) \\
    =\,& \gamma_{-}(t)^{-1} (u_0,g_0,u_0)\, \gamma_{-} (t)
\endaligned
$$
since, from the fact that $f \in I(G)$, we have $Df(g_0)=Ad_{g_0} Df(g_0)$.
Take
$$ (u(t),g(t),u(t)) =  \gamma_{+}(t)^{-1} (u_0,g_0,u_0)\, \gamma_{+} (t).$$
By differentiating the expression, we obtain
$$
\aligned
       & {d \over dt} (u(t),g(t),u(t)) \\
    =\,& T_{\epsilon (u(t))} {\bl}_{(u(t),g(t),u(t))}
         T_{\gamma_{+} (t)} {\bl}_{{\gamma_{+}(t)}^{-1}} {\dot\gamma_{+}(t)} 
         \qquad  \qquad \qquad (*)\\
       & + T_{\epsilon (u(t))} ({\br}_{(u(t),g(t),u(t))}\circ i)
           T_{\gamma_{+} (t)}{\bl}_{{\gamma_{+}(t)}^{-1}}{\dot \gamma_{-}(t)}.
\endaligned
$$
On the other hand, by rewriting (4.9) in the form
$$exp \{-t(0,Df(g_0),0)\}(u_0)\,\,{\gamma_{-}(t)} ={\gamma_{+}(t)},$$
we have
$$\eqalign{& T_{\gamma_{+} (t)} {\bl}_{{\gamma_{+}(t)}^{-1}} {\dot\gamma_{+}(t)}\cr
= \, & T_{\gamma_{-} (t)}{\bl}_{{\gamma_{-}(t)}^{-1}}{\dot \gamma_{-}(t)}\cr
  &-T_{{\gamma_{-}(t)}^{-1}} {\br}_{\gamma_{-}(t)} T_{\epsilon(u_{0})}
   {\bl}_{{\gamma_{-}(t)}^{-1}} (0_{u_{0}}, Df(g_0), 0).\cr}
$$
But 
$$
\aligned
       &T_{{\gamma_{-}(t)}^{-1}} {\br}_{\gamma_{-}(t)} T_{\epsilon(u_{0})}
        {\bl}_{{\gamma_{-}(t)}^{-1}} (0_{u_{0}}, Df(g_0), 0) \\
     =\, \, &(0_{u(t)}, Df(g(t)), 0),
\endaligned
$$
as $f\in I(G)$.  Hence it follows that

$$
\aligned
       &T_{\gamma_{+} (t)} {\bl}_{{\gamma_{+}(t)}^{-1}} {\dot\gamma_{+}(t)}
        -T_{\gamma_{-} (t)}{\bl}_{{\gamma_{-}(t)}^{-1}}{\dot \gamma_{-}(t)}\\
      =&-(0_{u(t)}, Df(g(t)), 0).
\endaligned
$$
From the property of $\gamma_{\pm}$ in (4.10), we can now conclude that

$$
\aligned
       &T_{\gamma_{\pm} (t)}{\bl}_{{\gamma_{\pm}(t)}^{-1}} {\dot \gamma_{\pm}(t)} \\
      =& -{1 \over 2} {\Cal R}_{u(t)}^{\pm} (0_{u(t)}, Df(g(t)),0).
\endaligned
$$
Therefore, on substituting into (*), we find
$$
\aligned
       & {d \over dt} (u(t),g(t),u(t)) \\
      =& - {1 \over 2} T_{\epsilon (u(t))} {\bl}_{(u(t),g(t),u(t))}
          {\Cal R}_{u(t)}^{+} (0_{u(t)}, Df(g(t)), 0) \\
       & -{1 \over 2} T_{\epsilon (u(t))} ({\br}_{(u(t),g(t),u(t))}\circ i)
          {\Cal R}_{u(t)}^{+}(0_{u(t)}, Df(g(t)), 0).
\endaligned
$$
Hence our claim follows from the fact that

$$
\aligned
       & T_{\epsilon (u(t))} {\bl}_{(u(t),g(t),u(t))} (0_{u(t)},Df(g(t)),0) \\
    =\,& -T_{\epsilon (u(t))} ({\br}_{(u(t),g(t),u(t))} \circ i)
          (0_{u(t)},Df(g(t)),0) \\
    =\,& (0_{u(t)}, T_e l_{g(t)} Df(g(t)), 0).
\endaligned
$$

To prove the existence of the factors $\gamma_{\pm}(t)$, we solve
the initial value problems
$$\dot k_{\pm}(t) = -{1 \over 2} T_e l_{k_{\pm}(t)} R^{\pm}(u(t)) Df(g(t)),
\qquad  k_{\pm}(0)=1, \qquad \quad (**)$$
where $u(t)$ and $g(t)$ are the solutions of (4.8) with initial
data $u(0) = u_0$, $g(0) = g_0$ (which are known
to exist by ODE theory).  Set 
$\gamma_{\pm} (t) = (u_0, k_{\pm}(t), u(t))$.  As can be easily 
verified, we can combine the equations for $u(t)$, $k_{\pm}(t)$ into
one single equation for $(\gamma_{+}(t),\gamma_{-}(t))$:
$$
\aligned
       &{d \over dt} (\gamma_{+}(t), \gamma_{-}(t)) \\
    =\,&\bigl(T_{\epsilon(u(t))} \bl_{\gamma_{+}(t)}
         {\Cal R}^{+} (0_{u(t)}, -{1\over 2} Df(g(t)), 0),
         T_{\epsilon(u(t))} \bl_{\gamma_{-}(t)}
         {\Cal R}^{-} (0_{u(t)}, -{1\over 2} Df(g(t)), 0)\bigr) \\
    =\,&T_{\epsilon_{P} (\beta_{P} (\gamma_{+}(t), \gamma_{-}(t)))}\,
        \bl^{P}_{(\gamma_{+}(t), \gamma_{-}(t))}
        ({\Cal R}^{+},{\Cal R}^{-}) (0_{u(t)}, -{1\over 2} Df(g(t)), 0)
     \quad \quad (***)
\endaligned
$$
where          
$\bl^{P}_{ (\gamma_{+}(t), \gamma_{-}(t))}$ represents left translation
by $(\gamma_{+} (t), \gamma_{-}(t))$ in the product groupoid
$P = \Gamma {\underset U\times U\to \times} \Gamma \rightrightarrows U$. 
Clearly, what we have just written down is a well-defined equation
for $(\gamma_{+}(t),\gamma_{-}(t))\in Im ({\Cal R}^{+},{\Cal R}^{-}).$
Moreover, from the initial conditions for $k_{\pm}(t)$ and $u(t)$, we
have $(\gamma_{+}(0),\gamma_{-}(0)) \in Im ({\Cal R}^{+},{\Cal R}^{-}).$

Now, from the equations for $k_{\pm}$ in (**),  we find
$$ 
\aligned
       &{d \over dt} \gamma_{+}(t)\, \gamma_{-}(t)^{-1}  \\
    =\,& (0_{u_0}, -T_{k_{+}(t)} r_{{k_{-}(t)}^{-1}} T_e l_{k_{+}(t)} Df(g(t)),
            0) \\
    =\,& (0_{u_0}, -T_e l_{k_{+}(t)k_{-}(t)^{-1}} Df (k_{-}(t) g(t)
            k_{-}(t)^{-1}), 0).
\endaligned
$$
But it follows on using the equations for $k_{\pm}$ and $g$ that
$$ {d \over dt} k_{-}(t) g(t) k_{-}(t)^{-1} = 0.$$
Therefore, $k_{-}(t) g(t) k_{-}(t)^{-1} = g_{0}$ and so
$$
\aligned
       &{d \over dt} \gamma_{+}(t)\, \gamma_{-}(t)^{-1}  \\
    =\,& (0_{u_0}, -T_e l_{k_{+}(t)k_{-}(t)^{-1}} Df (g_{0}), 0).
\endaligned
$$
As $\gamma_{+}(t)\,\gamma_{-}(t)^{-1} = (u_{0}, k_{+}(t)k_{-}(t)^{-1}, u_{0}),$
this shows that $k_{+}(t)k_{-}(t)^{-1} = e^{-t Df(g_{0})}$ and 
consequently, 
$$exp \{-t(0,Df(g_0),0)\}(u_0) = \gamma_{+}(t)\, \gamma_{-}(t)^{-1}.$$
Thus it remains to show that condition (4.10a) is satisfied.  But this
is immediate from (***).  This
completes the proof.
\pf
\enddemo

\proclaim
{Corollary 4.10} Let $\psi_{t}$ be the induced flow on ${\Cal I}\Gamma$ as
defined in (4.11) and let $\phi_{t}$ be the Hamiltonian flow of 
${\Cal F} =L^{*} f$ on $X$, where $L = Pr_{2}\circ\rho$ for a realization
map $\rho:X \longrightarrow \Gamma$ satisfying assumptions A1-A3.  If we can 
solve
for $\phi_{t} (x)$, $x \in J^{-1} (\mu)$ explicitly from the relation
$\rho(\phi_{t}) (x) = \psi_{t} (\rho (x))$, then the formula
$\phi^{red}_{t} \circ \pi_{\mu} = \pi_{\mu} \circ \phi_{t} \circ i_{\mu}$
gives an explicit expression for the flow of the reduced Hamitonian
${\Cal F}_{\mu} = {\widehat \rho}^{*} {\overline f}$.
\endproclaim
\smallskip
\noindent {\bf Remark 4.11} (a) The reader should not feel uneasy about the use
of equations (**) above (which involve the solutions $u(t)$ and $g(t)$)to 
show the existence of the factors $k_{\pm}(t)$, and which are then used
in turn to construct $u(t)$ and $g(t)$.  As the reader will see in 
Section 5, knowledge of the existence of the factorization facilitates 
its construction.
\smallskip
\noindent (b) For Hamiltonian systems which admit realization in the
dual vector bundles of coboundary dynamical Lie algebroids (where $R$ solves
(mDYBE)) and satisfy assumptions A1,A2 and A4 in \c{LX2}, there is also a 
method for solving the flows similar to what we discussed above.  We shall
refer the reader to \c{L2}, \c{L3} for details.
\smallskip
\noindent (c) Although we can apply Theorem 4.9 even when $R$ is a constant
r-matrix, it would be simpler to use the fact that the
vector bundles 
$\bigl \{{\Cal R}_q^{\pm} (0_q,X,0)\mid q\in U,\,X\in \fg \bigr \}$
are Lie subalgebroids of $A\Gamma$ in this case. An analog of 
Theorem 4.9 using these objects can be formulated for the constant
r-matrix case, but we provide no details here.
\bigskip
\bigskip

\subhead
5. \ A family of hyperbolic spin Ruijsenaars-Schneider models
\endsubhead

\bigskip
In this section, we shall construct
a family of hyperbolic spin Ruijsenaars-Schneider models using
coboundary dynamical Poisson groupoids.  To illustrate the factorization
method of Section 4, we shall solve the Hamilton's equations in a simple
case.
To do so, we shall make use of the solutions of (CDYBE) for
pairs $(\fg, \fh)$ of Lie algebras, as classified in \c{EV}.  Here, $\fg$
is simple, and $\fh \subset \fg$ is a Cartan subalgebra.

We begin with some notation.  Let
$\fg=\fh \oplus \sum_{\alpha \in\Delta} \fg_{\alpha}$ be
the root space decomposition of the simple Lie algebra
$\fg$ and let $(\cdot,\cdot)$ denote its Killing form.
We fix a simple system of roots  $\pi=\{\alpha_1,\cdots, \alpha_N\}$ 
and denote by
$\Delta^\pm$ 
the corresponding positive/negative
system.  For any positive root $\alpha \in \Delta^+$, we  choose
root vectors $e_{\alpha} \in \fg_{\alpha}$ and
$e_{-\alpha} \in \fg_{-\alpha}$ which are dual with respect to
$(\cdot, \cdot )$ so that $[e_{\alpha} , e_{-\alpha}]=h_{\alpha}$.  We
also fix an orthonormal basis $(x_i)_{1\le i\le N}$ of $\fh$.
Lastly, for a subset of simple roots $\pi^{\prime}\subset \pi$, we shall
denote the 
root span of $\pi^{\prime}$ by
$<\pi^{\prime}>\subset \Delta$ and set ${\overline
\pi^{\prime}}^{\pm}= \Delta^\pm \setminus <\pi^{\prime}>^\pm.$

For any subset $\pi^{\prime}\subset\pi$, we consider the following 
$H$-equivariant
solution of the (mDYBE) (with $K=
{1\over 2} id_{\fg}$):

$$R(q)X= - \sum_{\alpha\in \Delta} \phi_{\alpha} (q) X_{\alpha} e_{\alpha}
\eqno(5.1)$$
where
$$\eqalign {&\phi_\alpha (q) = {1\over 2}\,\, {\hbox { for }} \alpha\in
{\overline 
\pi^{\prime}}^+,\quad\phi_\alpha (q) = - {1\over 2} \,\,{\hbox { for }} 
\alpha\in
{\overline 
\pi^{\prime}}^-\cr
&\phi_\alpha (q)= {1\over 2} coth \bigl({1\over2} (\alpha(q))\bigr)\,\, 
{\hbox {for}} \alpha\in <\pi^{\prime}>,\cr}$$
and $X_{\alpha}= (X, e_{-\alpha}), \quad \alpha\in \Delta.$

Throughout this section, the coboundary dynamical Poisson groupoid
$(\Gamma, \{\cdot, \cdot \}_R)$ will refer to the one which corresponds
to this choice of $R$. Also, we assume the Lie groups $G$ and $H$ are
simply-connected.

Let $\omega_1,\ldots, \omega_N$ be fundamental weights of $\fg$, and
let $\chi_1,\ldots, \chi_N$ denote the characters of the irreducible
representations corresponding to these fundamental weights \c{St}.

\definition
{Definition 5.1} The spin Ruijsenaars-Schneider models associated
to $R$ are the Hamiltonian systems on $\Gamma$ generated by
nonzero multiples of $H_i = Pr_2^{*}\chi_i$, $i = 1,\cdots, N.$
\enddefinition

\example
{Example 5.2} Take $G=SL(N+1, \Bbb C)$ and let $H$ be the Cartan
subgroup consisting of diagonal matrices.  We shall denote the
corresponding Lie algebra by $\fg$ and $\fh$ respectively and we
shall use the pairing $(A,B)=tr (AB)$ on $\fg$.
Consider the spin Ruijsenaars-Schneider model generated by 
$H_1(q, g, q^{\prime}) = \chi_1 (g) = tr (g)$ for
the case where $\pi^{\prime} = \pi$.  Since on ${\Cal I} \Gamma$, 
the Hamilton's
equation is of the form $\dot q= -{1 \over 2} \Pi_{\fh}\,(g-{1\over N+1}
tr (g) I)$, 
$\dot g = {1 \over 2} (R(q) g) g - {1 \over 2} g (R(q) g),$ it follows
that in terms of the components of $q$ and $g$, we have 
$$\ddot q_{i} = -{1\over 2} \dot g_{ii}
  = {1\over 4}\sum_{k\neq i} \coth ((q_{i}-q_{k})/2) g_{ik} g_{ki},
  \eqno(5.2a)$$
$$\eqalign{\dot g_{ij} =& {1 \over 4} \coth ((q_{i}-q_{j})/2) g_{ij}
    (g_{ii}-g_{jj})\cr
    & + {1\over 4} \sum_{k\neq i,j} \bigl( \coth((q_{k}-q_{j})/2)
    - \coth((q_{i}-q_{k})/2)\bigr) g_{ik} g_{kj}, \quad i\neq j\cr}
   \eqno(5.2b)
$$
Thus up to constants, these are exactly Eqns.(14)-(15) in the paper
 \c{BH} (compare also Eqns.(1.21)-(1.23) of \c{KZ}) if we take $q$
to be real diagonal and $g$ to be Hermitian.  It is a remarkable
fact that these are the equations which govern the soliton dynamics of
the so-called $A^{(1)}_N$ affine Toda field theory \c{BH}.
\endexample

Rather than spelling out the equations of the other systems explicitly
in terms of components, our next goal is to solve the equations via
the factorization method in Section 4.  For simplicity, we shall do
it for the special case where $\pi^{\prime} = \pi$, the lengthy analysis
of the case where $\pi^{\prime}\neq \pi$ is given in \c{L2}.

From the definition of $R$ in (5.1), it is straightforward to
check that in this case, the bundle maps ${\Cal R}^{\pm}$ are 
isomorphisms of Lie algebroids so that $Im {\Cal R}^{\pm} = A\Gamma.$
To set up the factorization problem, it is important to give a
precise description of $Im ({\Cal R}^{+}, {\Cal R}^{-}) \subset
A\Gamma {\underset TU \to \oplus} A\Gamma$.  In this connection, note
that the bundle map
$$\theta : Im {\Cal R}^{+} \longrightarrow Im {\Cal R}^{-}:
   {\Cal R}^{+} (0_q, X, Z)\mapsto {\Cal R}^{-} (0_q, X, Z)\eqno(5.3)
$$
is well-defined and moreover is a Lie algebroid isomorphism (note that
this is not so if $\pi^{\prime}\neq \pi$).  Indeed, we have
$\theta (0_q, X, Z)= (0_q, -\iota Z + Ad_{e^q} X, Z)$.  Therefore,
$$\eqalign{& Im ({\Cal R}^{+}, {\Cal R}^{-})\cr
           =&\bigl \{((0_q, X, Z), \theta (0_q, X, Z))\mid q\in U,
            X\in \fg, Z\in \fh \bigr \}.\cr}\eqno(5.4)
$$
Now, we can check that $\theta$ integrates to an isomorphism of the
Lie groupoid $\Gamma$, given by $(u, g, v)\mapsto (u, e^{u} g e^{-v}, v)$.
Hence the factorization problem in (4.9) reduces to 
$$e^{-tDf(g_{0})}e^{u_0} = k_{+}(t) e^{u(t)} {k_{+}(t)}^{-1}. \eqno(5.5)$$
As the union of the conjugates of $H$ forms an open dense subset of 
$G$, we can find (for at least small values of $t$) $x_{+}(t)\in G$ ($x_{+}(t)$
is unique up to transformations $x_{+}(t)\rightarrow x_{+}(t) \delta (t)$, 
where $\delta (t) \in H$)
and unique $d(t)\in H$ such that
$$e^{-tDf(g_{0})}e^{u_0} = x_{+}(t) d(t) {x_{+}(t)}^{-1}\eqno(5.6)$$
with $x_{+}(0)=1, d(0)= e^{u_0}$.  This uniquely determines $u(t)$ via
the formula
$$u(t) = log \, d(t).\eqno(5.7)$$
On the other hand, let us fix one such $x_{+}(t)$.  We shall seek
$k_{+}(t)$ in the form 
$$k_{+}(t) = x_{+}(t) h(t), \quad h(t) \in H.\eqno(5.8)$$
To determine $h(t)$, we shall impose the condition in Eqn. (4.10).
When we write out this condition, we find it natural to introduce
$g_{+}(t) = k_{+} (t) e^{-{1\over 2} (u(t)-u_{0})}$, in terms of
which the condition becomes
$$\Pi_{\fh} \bigl (T_{g_{+}(t)} l_{g_{+}(t)^{-1}} \dot g_{+}(t) \bigr ) =0.
\eqno(5.9)$$
From this, we find that $h(t)$ satisfies the equation
$$\dot h(t) = T_{e} l_{h(t)}\bigl ({1\over 2} \dot u(t) - \Pi_{\fh} 
(T_{x_{+}(t)} l_{{x_{+}(t)}^{-1}} \dot x_{+}(t)) \bigr )\eqno(5.10)$$
with $h(0)=1$.  Solving this equation explicitly, we find that
$$k_{+}(t) = x_{+}(t) exp\{{1\over 2}(u(t) - u_{0})-\int _{0}^{t}{\Pi_{\fh} 
(T_{x_{+}(\tau)} l_{{x_{+}(\tau)}^{-1}} 
\dot x_{+}(\tau))}\, d\tau \} .\eqno(5.11)$$
Hence we we can write down the solution using (4.11).

\enddocument